\documentclass[journal]{IEEEtran}

\usepackage{amsmath,graphicx}
\newtheorem{theorem}{\bf Theorem}
\newtheorem{proposition}{\bf Proposition}

\newtheorem{corollary}{\bf Corollary}

\begin{document}
\title{Squeezing the Arimoto-Blahut algorithm for faster convergence}

\author{Yaming~Yu, {\it Member, IEEE}
\thanks{Yaming Yu is with the Department of Statistics, University of California, Irvine, CA, 92697-1250, USA
(e-mail: yamingy@uci.edu).  This work is supported in part by a start-up fund from the Bren School of Information and 
Computer Sciences at UC Irvine.}
}

\maketitle

\begin{abstract}
The Arimoto--Blahut algorithm for computing the capacity of a discrete memoryless channel is revisited.  A so-called  ``squeezing'' strategy is used to design algorithms that preserve its simplicity and monotonic convergence properties, but have provably better rates of convergence. 
\end{abstract}

\begin{IEEEkeywords}
alternating minimization; channel capacity; discrete memoryless channel; rate of convergence.
\end{IEEEkeywords}

\section{Introduction}
The Arimoto--Blahut Algorithm \cite{A, B} (ABA) plays a fundamental role in numerical calculations of channel capacities.  This iterative scheme has an appealing geometric interpretation (\cite{CT}), and possesses a desirable monotonic convergence property.  We refer to \cite{N, V, RG, DYW, MD} for extensions and improvements. 

We study variants of ABA with an aim to speed up the convergence while maintaining the simplicity.  The focus is on the discrete memoryless channel, and on theoretical properties; extensions and further numerical results will be reported in future works.  Our investigation relies on certain reformulations that slightly generalize the original capacity calculation problem.  Each formulation leads to an Arimoto-Blahut-type algorithm, which is monotonically convergent, and typically as easily implemented as the original ABA.  A formula for the rate of convergence provides valuable insight as to when ABA is slow.  Comparison theorems show that our constructions are at least as fast as the usual ABA as measured by the global convergence rate.  Numerical examples show that the improvement can be substantial. 

Our approach differs from other acceleration methods for ABA (e.g., the proximal point formulation of \cite{MD}) in that we focus on preprocessing or ``reparameterizing'' the problem (Sections III and IV).  Such reparameterizations, broadly termed ``squeezing,'' aim at reducing the overlap between rows of the channel matrix.  Our technical contributions include the monotonic convergence theorem of Section IV, and the convergence rate comparison theorems of Section V.  These theoretical results are illustrated with simple examples. 

\section{Variants of Arimoto-Blahut}
A discrete memoryless channel is associated with an $m\times n$ transition matrix $W=(W_{ij})$, i.e., $W_{ij}$ specifies the probability of receiving the output letter $j$ if the input is $i$.  Mathematically $W_{ij}\geq 0$ and $\sum_j W_{ij}=1$ for all $i$.  The information capacity is defined as 
\begin{equation}
\label{ip}
\sup_{p\in \Omega} I(p),\quad I(p) =\sum_i p_i D(W_i||pW).
\end{equation}
See \cite{G, Cover} for interpretations of this fundamental quantity.  Throughout $\Omega$ denotes the probability simplex 
$$\Omega=\{p=(p_1, \ldots, p_m):\ p_i\geq 0,\ p 1_m=1 \},$$
$1_m$ denotes the $m\times 1$ vector of ones, $W_i$ denotes the $i$th row of $W$, i.e., $W_i=(W_{i1},\ldots, W_{i n})$, and $D(q||r)=\sum_i q_i\log (q_i/r_i)$ for nonnegative vectors $q=(q_i)$ and $r=(r_i)$.  We use natural logarithm (except for Fig.\ 3) and obey the convention $0\log (0/a)=0,\ a\geq 0$.  Let us also define $H(q)=-\sum_i q_i\log q_i$ for a nonnegative vector $q=(q_i)$.  It is not required that $\sum_i q_i=1$.  Without loss of generality assume that not all rows of $W$ are equal, and that none of its columns is identically zero. 

An example of our general class of algorithms for solving (\ref{ip}) is as follows.  Let $\lambda\in \mathbf{R}$ satisfy 
\begin{equation}
\label{lam}
1\leq \lambda \leq \frac{1}{1-\sum_j \min_i W_{ij}}.
\end{equation}
{\bf Algorithm I: Singly Squeezed ABA.}
Choose $p^{(0)}\in \Omega$ such that $p^{(0)}_i>0$ for all $i$.  For $t=0, 1, \ldots$, calculate $p^{(t+1)}$ as 
\begin{equation}
\label{alg1}
p_i^{(t+1)}=\frac{p_i^{(t)}\exp\left(\lambda z^{(t)}_i\right)}{\sum_l p_l^{(t)} \exp\left(\lambda z_l^{(t)}\right)};\quad z_i^{(t)}=D\left(W_i||p^{(t)}W\right).
\end{equation}
Iterate until convergence. 

One recognizes Algorithm I as a generalization of the original Arimoto-Blahut Algorithm, which corresponds to $\lambda=1$.  This simple generalization has been considered before (see, e.g., \cite{MD}).  What is new is the constraint (\ref{lam}).  Under this constraint, Algorithm I is guaranteed to converge monotonically (Section IV), and its convergence rate is no worse than that of ABA (Section V).  The nickname reflects our intuitive interpretation of Algorithm I and is explained near the end of Section III. 

{\bf Example 1.}
Consider the channel matrix 
$$W=\left(\begin{array}{ccc} 0.7 & 0.2 & 0.1\\ 0.1 & 0.2 & 0.7
\end{array}\right)$$
which is also used by \cite{MD} as an illustration.  Let us choose $\lambda=5/3$, which attains the upper bound in (\ref{lam}).  Fig.\ 1 compares the iterations $p^{(t)}_1,\ t=1, 2,\ldots,$ produced by ABA and by Algorithm I with $\lambda=5/3$.  Each algorithm is started at $p^{(0)}=(1/3,\, 2/3)$.  Algorithm I, however, appears to approach the target $p^*=(1/2,\, 1/2)$ faster than ABA.  Different starting values give similar comparisons. 
\begin{figure}
\begin{center}
\includegraphics[width=2.2in, height=3.4in, angle=270]{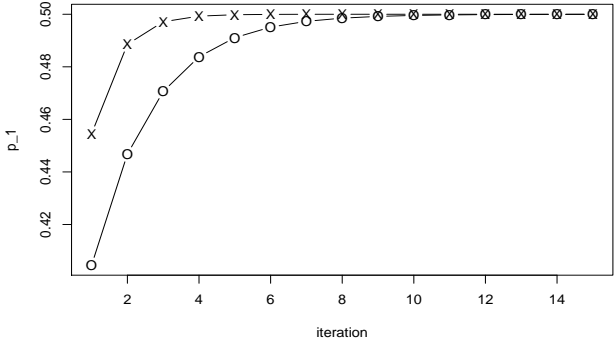}
\end{center}
\caption{Iterations of $p_1^{(t)}$ for ABA (``--O--'') and Algorithm I (``--X--'') with $\lambda=5/3$.}
\end{figure}

Algorithm I is a special case of the following class of algorithms.  Henceforth define 
$$\Omega(r)=\{p\in \Omega:\ p\geq r\}$$
for any $1\times m$ vector $r\geq 0$.  For vectors (matrices) $A$ and $B$ of the same dimension, $A\geq B$ means every entry of $A-B$ is nonnegative.  The $m\times m$ identity matrix is written as $I_m$. 

Let $r$ be a nonnegative $1\times m$ vector such that 
\begin{align}
\label{W}
   &W\geq 1_m r W.
\end{align}
Define $r_+=r 1_m$.  Let $\lambda$ (a scalar) satisfy 
\begin{equation}
\label{lam2}
\frac{1}{1-r_+}\leq \lambda\leq \frac{1}{1-\sum_j \min_i W_{ij}}.
\end{equation}

{\bf Algorithm II: Doubly Squeezed ABA.}
Choose $p^{(0)}\in \Omega(r)$ such that $p^{(0)}_i>0$ for all $i$.  For $t=0, 1, \ldots$, calculate 
\begin{equation}
\label{water}
p_i^{(t+1)} =\max\left\{r_i,\ \delta^{(t)} p^{(t)}_i \exp\left(\lambda z_i^{(t)}\right)\right\}
\end{equation}
where 
$$z_i^{(t)}=D\left(W_i||q^{(t)}W\right),\quad q^{(t)}=\frac{p^{(t)}-r}{1-r_+},$$ 
and $\delta^{(t)}$ is chosen such that $\sum_i p_i^{(t+1)}=1$.  Upon convergence, output 
$$\hat{p}=\frac{p^{(\infty)}-r}{1-r_+}.$$ 

A stopping criterion for practical implementation is ($\epsilon>0$)
\begin{equation}
\label{converge}
\max_i z_i^{(t)} -\sum_i q^{(t)}_i z_i^{(t)}\leq \epsilon.
\end{equation}
This is the same criterion as often used for ABA (\cite{B}), and it is convenient since the quantities $z_i^{(t)}$ are readily available at each iteration. 

A key requirement is (\ref{W}).  It implies, for example, 
$$r_+ \leq \sum_j \min_i W_{ij}<1,$$
assuming that not all rows of $W$ are equal.  When $m=2$, (\ref{W}) becomes 
\begin{align}
\label{r1}
\frac{r_1}{1-r_1-r_2} &\leq \min_{j:\ W_{1j}>W_{2j}} \frac{W_{2j}}{W_{1j}-W_{2j}},\quad {\rm and}\\
\label{r2}
\frac{r_2}{1-r_1-r_2} &\leq \min_{j:\ W_{2j}>W_{1j}} \frac{W_{1j}}{W_{2j}-W_{1j}}.
\end{align}
For general $m$, the restrictions on $r$ are less clear.  See Section V for further discussion. 

If $r\equiv 0$, then (\ref{water}) reduces (\ref{alg1}), showing Algorithm II as a generalization of Algorithm I.  Compared with Algorithm I, Algorithm II is only slightly more difficult to implement.  In (\ref{water}), determining $\delta^{(t)}$ is a form of waterfilling (\cite{Cover}), which can be implemented in $O(m\log m)$ time.  (A simple implementation is included in Appendix A for completeness.)  Hence the additional cost per iteration is minor.  The improvement in convergence rate, however, can be substantial. 

{\bf Example 1 (continued).} Consider Algorithm II with $\lambda=5/3$ and $r=(1/8,\, 1/8)$.  Then (\ref{r1}) and (\ref{r2}) are satisfied with equalities.  Inspection of (\ref{water}) reveals that we have $p^{(1)}=(1/2,\, 1/2)$, regardless of the starting value $p^{(0)}$. (It is easier to verify this with the equivalent form of Algorithm II in Section III.)  That is, with this choice of $\lambda$ and $r$, Algorithm II converges in one step. 

The general validity of Algorithm II is verified in Section IV.  The critical issue of which values of $r$ and $\lambda$ lead to fast convergence is studied in Section V, where theoretical justifications are provided for the following guideline.  For fast convergence, we should 
\begin{itemize}
\item
set $\lambda$ at the upper bound in (\ref{lam2}), and 
\item
let $r/(1-r_+)$ be as large as possible, subject to the restriction (\ref{W}). 
\end{itemize}
For $m=2$, this means that $r$ should satisfy the equalities in (\ref{r1}) and (\ref{r2}).  Although Example 1 already hints at such a recommendation, we also conduct a simulation for illustration. 

{\bf Example 2.}  A channel matrix $W$ with $m=2$ and $n=8$ is generated according to $W_{ij}=u_{ij}/\sum_k u_{ik}$ where $u_{ij}$ are independent uniform($0,1$) variates.  The original ABA, Algorithm I, and Algorithm II are compared.  For Algorithm I, we set $\lambda$ at the upper bound in (\ref{lam}); for Algorithm II, we choose $r/(1-r_+)$ to satisfy the upper bounds in (\ref{r1})--(\ref{r2}), and set $\lambda$ at the upper bound in (\ref{lam2}).  The starting values are $p^{(0)}=(1/2, 1/2)$ for ABA and Algorithm I, and $p^{(0)}=(1-r_+)(1/2, 1/2)+r$ for Algorithm II.  We record the number of iterations until the common criterion (\ref{converge}) is met with $\epsilon=10^{-8}$.  The experiment is replicated 100 times. 

The improvement in speed by using Algorithm I or Algorithm II is evident from Fig.\ 2, which displays two bivariate plots of the numbers of iterations.  While ABA sometimes takes hundreds of iterations, Algorithm I takes no more than $40$, and Algorithm II no more than $16$, throughout the 100 replications.  The large reduction in the number of iterations is also shown in Fig.\ 3, which summarizes the $\log_2$ acceleration ratios, defined as $\log_2(N_{ABA}/N_I)$ for Algorithm I, for example.  Here $N_{ABA}$ (resp.\ $N_I$) denotes the number of iterations for ABA (resp.\ Algorithm I). The median acceleration ratio is $4.0$ for Algorithm I, and around $7.1$ ($2^{2.83}$) for Algorithm II.  The minimum acceleration ratio is $2.2$ for Algorithm I and $2.8$ for Algorithm II.  Overall this supports the preference for large values of $\lambda$ and $r/(1-r_+)$, subject to (\ref{W}) and (\ref{lam2}), in implementing Algorithm II. 

\begin{figure}
\begin{center}
\includegraphics[width=2.2in, height=3.4in, angle=270]{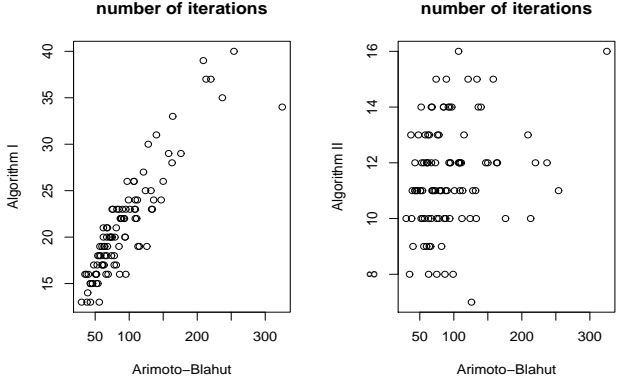}
\end{center}
\caption{Comparing the numbers of iterations for three algorithms in Example 2.}
\end{figure}

\begin{figure}
\begin{center}
\includegraphics[width=2.2in, height=3.4in, angle=270]{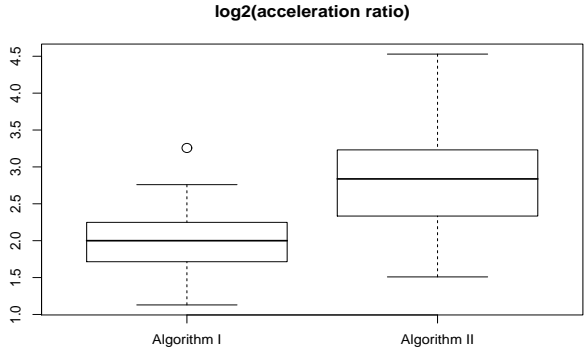}
\end{center}
\caption{Acceleration ratios in Example 2.}
\end{figure}

{\bf Remark.}  One may still implement Algorithm II with some $r,\ \lambda$ that do not satisfy (\ref{W}) or (\ref{lam2}).  For example, it is conceivable that values of $\lambda$ slightly exceeding the upper bound in (\ref{lam2}) could lead to even faster convergence.  However, our theoretical results only guarantee convergence under (\ref{W}) and (\ref{lam2}).  It is also intuitive that setting $\lambda$ too large would overshoot and no longer maintain monotonic convergence. 

\section{Equivalent form of Algorithm II}
Although Algorithm II is convenient for practical implementation, we write it in an equivalent form (Algorithm III) to study the theoretical properties. 

Let $r$ ($1\times m$) and $f$ ($1\times n$) be nonnegative vectors that satisfy 
\begin{align}
\label{tW}
\tilde{W} \equiv (1+f_+)\frac{I_m-1_m r}{1-r_+} W -1_m f \geq 0,\quad r_+\equiv r 1_m <1,
\end{align}
and $f_+\equiv f 1_n$.  Set
\begin{equation}
\label{c}
c_i =H(\tilde{W}_i)-\frac{1+f_+}{1-r_+}H(W_i),\quad 1\leq i\leq m.
\end{equation}

{\bf Algorithm III: Doubly Squeezed ABA.}
Choose $p^{(0)}\in \Omega(r)$ such that $p^{(0)}_i>0$ for all $i$.  For $t=0, 1, \ldots$, calculate 
\begin{align}
\label{phit}
\Phi^{(t)}_{ji} &=\frac{p^{(t)}_i \tilde{W}_{ij}}{f_j +\sum_l p^{(t)}_l\tilde{W}_{lj}};\\
\label{pt+1}
p_i^{(t+1)} &=\max\left\{r_i,\ \alpha^{(t)} e^{c_i+\sum_j \tilde{W}_{ij}\log \Phi^{(t)}_{ji}}\right\},
\end{align}
where $\alpha^{(t)}$ is chosen such that $\sum_i p_i^{(t+1)}=1$.  Upon convergence, output 
$$\hat{p}=\frac{p^{(\infty)}-r}{1-r_+}.$$ 

The restriction (\ref{tW}) can be broken down as
\begin{equation}
\label{rsat}
r_+<1,\quad W^*\equiv \frac{I_m-1_m r}{1-r_+}W\geq 0,
\end{equation}
and
\begin{equation}
\label{fsat}
(1+f_+)W^*-1_m f\geq 0.
\end{equation}
The restriction (\ref{rsat}) is a restatement of (\ref{W}), while (\ref{fsat}) is equivalent to
\begin{equation}
\label{W*}
f_j\leq (1+f_+)\min_i W^*_{ij},\quad j=1,\ldots, n.
\end{equation}

If we set 
\begin{equation}
\label{lamfr}
\lambda=\frac{1+f_+}{1-r_+},
\end{equation}
then Algorithm III reduces to Algorithm II.  Indeed, by summing over $j$, (\ref{W*}) leads to
$$f_+\leq \frac{1+f_+}{1-r_+} \sum_j \left[\min_i W_{ij} -(rW)_j\right],$$
from which we obtain the upper bound in (\ref{lam2}).  Moreover, after some algebra, the mapping $p^{(t)}\to p^{(t+1)}$ as specified by (\ref{phit})--(\ref{pt+1}) reduces to (\ref{water}).  (A useful identity in this calculation is $p\tilde{W}+f=\lambda (p-r)W$; see also Proposition \ref{prop1} in Section IV.)  Thus Algorithm III reduces to Algorithm II with $\lambda$ given by (\ref{lamfr}).

Conversely, suppose $r$ and $\lambda$ satisfy (\ref{W}) and (\ref{lam2}).  If we define 
$$f_j=[\lambda(1-r_+)-1]\frac{\min_i W^*_{ij}}{\sum_k \min_i W^*_{ik}},$$
with $W^*$ given by (\ref{rsat}), then (\ref{lamfr}) is satisfied.  We also deduce $f_j\geq 0$ and (\ref{W*}) from (\ref{lam2}).  Thus Algorithm II is equivalent to Algorithm III with this choice of $f$. 

We shall show that Algorithm II/III converges monotonically, and its convergence rate is no worse than that of ABA.  Intuitively, ABA is slow when there exists a heavy overlap between rows of the channel matrix $W$.  Algorithm III, which works with $\tilde{W}$ rather than $W$, can be seen as trying to reduce this overlap.  Its nickname is derived from the transformation (\ref{tW}), which subtracts, or ``squeezes out,'' a nonnegative vector from each row of $W$.  If $r\equiv 0$, then only a vector proportional to $f$ is subtracted.  But Algorithm III with $r\equiv 0$ is equivalent to Algorithm II with $r\equiv 0$, which is simply Algorithm I.  Hence Algorithm I is called ``Singly Squeezed ABA".  For general $r$ and $f$, we squeeze out both a vector proportional to $f$ and another one proportional to $rW$.  Hence Algorithm II/III is called ``Doubly Squeezed ABA".  The vector $r$ also modifies the space $\Omega$ we work on, thus making $rW$ separate from $f$. 

{\bf Example 1 (continued).} Consider Algorithm III with $r=(1/8,\, 1/8)$ and $f=(0,\, 1/4,\, 0)$.  This corresponds to Algorithm II with the same $r$ and $\lambda=5/3$.  By (\ref{tW}) 
we have 
$$\tilde{W}=\left(\begin{array}{ccc} 1 & 0 & 0\\ 0 & 0 & 1
\end{array}\right).$$
The rows of $\tilde{W}$ no longer overlap, i.e., $\tilde{W}_{1j}\tilde{W}_{2j}=0$ for all $j$.  Inspection of (\ref{phit}) and (\ref{pt+1}) reveals that we have $\Phi^{(1)}=\tilde{W}^\top$ and $p^{(1)}=(1/2,\, 1/2)$, regardless of the starting value $p^{(0)}$.  Thus, as mentioned earlier, Algorithm II/III converges in one step.

\section{Validity of Algorithm II/III}
Given an $m\times n$ stochastic matrix $V$, a $1\times n$ vector $f\geq 0$, a $1\times m$ vector $c$, and $p\in \Omega$, let us define 
$$I(p|V, f, c)=\sum_i p_i (D(V_i||f+pV)+c_i)+D(f||f+pV). $$
Equivalently, 
\begin{equation}
\label{ent}
I(p|V, f, c)=H(pV+f)+\sum_i p_i (c_i-H(V_i))-H(f).
\end{equation}

We have $I(p|W, 0, 0)=I(p)$ as in (\ref{ip}).  However, there exist less obvious relations.  Proposition \ref{prop1} is key to our derivation of Algorithm III.
\begin{proposition}
\label{prop1}
Let $r,\ f,\ \tilde{W},$ and $c=(c_1, \ldots, c_m)$ satisfy (\ref{tW}) and (\ref{c}).  Then
\begin{align}
\label{lin2}
I(p|W, 0, 0)= &\frac{I(\tilde{p}|\tilde{W}, f, c)+H(f)}{1+f_+}\\
\nonumber
&+\log(1+f_+)+ \frac{\sum_i r_i H(W_i)}{1-r_+},
\end{align}
where $\tilde{p}=(1-r_+)p+r.$
\end{proposition}
\begin{IEEEproof}
Noting 
$$\tilde{p}\tilde{W}+f=(1+f_+)pW,$$
the claim follows from (\ref{ent}) and routine calculations. 
\end{IEEEproof}

Relation (\ref{lin2}) implies that, in order to maximize $I(p|W, 0, 0)$ over $p\in \Omega$, we may equivalently maximize $I(\tilde{p}|\tilde{W}, f, c)$ over $\tilde{p}\in \Omega(r)$, and then set $p=(\tilde{p}-r)/(1-r_+)$.  Let us consider solving this slightly more general problem. 

{\bf Problem I.} Let $\tilde{W}$ be an $m\times n$ stochastic matrix, let $f\geq 0$ be a $1\times n$ vector, and let $r,\ c$ be $1\times m$ vectors.  Assume $r\geq 0$ and $r_+\equiv r 1_m<1$.  Maximize $I(p|\tilde{W}, f, c)$ over $p\in \Omega(r).$ 

Problem I can be handled by a straightforward extension of ABA.  Following \cite{A, B}, we note that maximizing $I(p|\tilde{W}, f, c)$ is equivalent to maximizing 
$$I(p, \Phi)=
\sum_{i\geq 1, j} p_i \tilde{W}_{ij}\log\frac{\Phi_{ji}}{p_i} + \sum_j f_j \log \Phi_{j0} + \sum_{i\geq 1} c_i p_i
$$
over $p\in \Omega(r)$ and $\Phi$ (an $n\times (m+1)$ stochastic matrix) jointly.  This holds because, for fixed $p,\ I(p, \Phi)$ is maximized by 
\begin{equation}
\label{maxphi}
\Phi_{ji}=\frac{p_i \tilde{W}_{ij}}{f_j +\sum_l p_l\tilde{W}_{lj}},\quad \Phi_{j0}=\frac{f_j}{f_j+\sum_l p_l \tilde{W}_{lj}},
\end{equation}
and the maximum value is $I(p|\tilde{W}, f, c)$.  On the other hand, for fixed $\Phi,\ I(p, \Phi)$ is maximized by 
\begin{equation}
\label{maxp1}
p_i=\max\left\{r_i,\ \alpha e^{c_i+\sum_j \tilde{W}_{ij}\log \Phi_{ji}} \right\},
\end{equation}
where $\alpha$ is chosen such that $\sum_i p_i=1$.  This verifies the Karush-Kuhn-Tucker conditions.  If $r\equiv 0$, then (\ref{maxp1}) reduces to 
\begin{equation}
\label{maxp}
p_i=\frac{\exp(c_i+\sum_j \tilde{W}_{ij}\log \Phi_{ji})}{\sum_{l\geq 1} \exp(c_l+\sum_j \tilde{W}_{lj}\log \Phi_{jl})}.
\end{equation}

Algorithm III simply alternates between (\ref{maxphi}) and (\ref{maxp1}).  At each iteration, the function $I(p|\tilde{W}, f, c)$ never decreases.  Theorem \ref{thm1} shows that Algorithm III converges to a global maximum.  The proof uses the alternating minimization interpretation of \cite{CT}; see Appendix B. 

\begin{theorem}[monotonic convergence]
\label{thm1}
Let $p^{(t)}$ be a sequence generated by Algorithm III.  Then $\lim_{t\to\infty} p^{(t)}\equiv p^{(\infty)}$
exists and, as $t\nearrow \infty$, 
$$I(p^{(t)}|\tilde{W}, f, c)\nearrow \sup_{p\in \Omega(r)} I(p|\tilde{W}, f, c).$$  
\end{theorem}

By Proposition \ref{prop1}, $(p^{(\infty)}-r)/(1-r_+)$ is a global maximizer of $I(p|W, 0, 0)$ over $p\in \Omega$.  That is, Algorithm III correctly solves the optimization problem (\ref{ip}) in the limit. 

\section{Rate of convergence} 
Throughout this section the notation of Algorithm III is assumed.  For example, $\tilde{W}$ is defined via (\ref{tW}).  
We derive a general formula (Theorem \ref{thm2}) for the rate of convergence.  Comparison results (Theorems \ref{thm3} and \ref{thm4}) show that Algorithm III is at least as fast the original ABA.  Based on the comparison theorems, a general recommendation is to let $r$ and $f$ (``the squeezing parameters'') be as large as permitted for fast convergence. 

Assume the iteration (\ref{phit})--(\ref{pt+1}) converges to some $p^*$ in the interior of $\Omega(r)$, i.e., $p^*_i>r_i$ for all $i$.  Denote the mapping from $p^{(t)}\to p^{(t+1)}$ by $M$.  Then $p^*=M(p^*)$, i.e., $p^*$ is a fixed point.  We emphasize that, because $p^*$ is assumed to lie in the interior of $\Omega(r)$, so are all $p^{(t)}$ for large enough $t$.  Hence (\ref{pt+1}) eventually takes the form of (\ref{maxp}), i.e., 
\begin{equation*}
p^{(t+1)}_i=\frac{\exp(c_i+\sum_j \tilde{W}_{ij}\log \Phi^{(t)}_{ji})}{\sum_{l\geq 1} \exp(c_l+\sum_j \tilde{W}_{lj}\log \Phi^{(t)}_{jl})}.
\end{equation*}

We call $R(p^*)=\partial M(p^*)/\partial p$ the ($m\times m$) {\it matrix rate of convergence} of Algorithm III, because
$$p^{(t+1)}-p^*\approx (p^{(t)}-p^*) R(p^*)$$
for $p^{(t)}$ near $p^*$.  The spectral radius of $R(p^*)$, written as $S(R(p^*))$, is called the {\it global rate of convergence}.  (The smaller the rate, the faster the convergence.)  Such notions are not uncommon in analyzing fixed point algorithms (see, e.g., \cite{DLR} and \cite{XLM}).  Technically, the global rate should be defined as the spectral radius of a restricted version of $R(p^*)$, because $(p^{(t)}-p^*) 1_m=0$.  However, the spectral radius of $R(p^*)$ is the same without this restriction (see Appendix D). 

The matrix $R(p^*)$ admits a simple formula (Theorem \ref{thm2}); see Appendix C for its proof. 
\begin{theorem}[rate of convergence]
\label{thm2} 
We have 
\begin{equation}
\label{key}
R(p^*) =I_m -\tilde{W}\Psi,
\end{equation}
where the $n\times m$ matrix $\Psi=(\Psi_{ji})$ is specified by 
\begin{equation}
\label{Psi}
\Psi_{ji}=\Phi_{ji}(p^*)+p_i^*\Phi_{j0}(p^*),\quad 1\leq j\leq n,\ 1\leq i\leq m,
\end{equation}
and $\Phi_{ji}(p^*)$ is $\Phi_{ji}$ as in (\ref{maxphi}) when taking $p=p^*$. 
\end{theorem}

For the original ABA, we have $R(p^*)=I_m-W\Phi(p^*)$, which can be broadly interpreted as a measure of how noisy the channel is.  If $m=n$ and $W$ approaches $I_m$, then so does $\Phi(p^*)$, and $R(p^*)$ approaches zero.  At the opposite end, if rows of $W$ overlap almost entirely, then $W\Phi(p^*)$ is nearly singular, leading to a large $S(R(p^*))$, and slow convergence for ABA.  See Corollary \ref{coro1} for a more quantitative statement. 

{\bf Example 1 (continued).} The maximizer of $I(p)$ is $\hat{p}=(1/2,\, 1/2)$.  The matrix rates are calculated for ABA ($R_0$) and for Algorithm III with $r\equiv 0$ and $f=(1/6,\, 1/3,\, 1/6)$ ($R_1$), which is equivalent to Algorithm I with $\lambda=5/3$: 
$$R_0=0.275\left(\begin{array}{cc} 1 & -1\\ -1 & 1\end{array}\right);\quad R_1=0.125\left(\begin{array}{cc} 1 & -1\\ -1 & 1\end{array}\right).$$
The global rates are $S(R_0)=0.55$ and $S(R_1)=0.25$.  Thus we confirm the advantage of this choice of $\lambda$ for Algorithm I.  For Algorithm III with $r=(1/8,\, 1/8)$ and $f=(0,\, 1/4,\, 0)$, the global rate is zero. 

Propositions \ref{prop2} and \ref{prop3} explore basic properties of $R(p^*)$; see Appendix D for the proofs.
\begin{proposition}
\label{prop2}
We have
\begin{itemize}
\item[1]
$R(p^*) 1_m=0$;
\item[2]
if $f\equiv 0$, then $R(p^*)$ is diagonalizable.
\end{itemize}
\end{proposition}

\begin{proposition}
\label{prop3}
If $d$ is an eigenvalue of $R(p^*)$, then $d$ is real and $0\leq d\leq 1$.
\end{proposition} 

Propositions \ref{prop2} and \ref{prop3} are used in deriving our main comparison results for convergence rates.  Let us write the {\it global rate} for Algorithm III as $R(r, f)$ to highlight its dependence on the vectors $r$ and $f$.  The global rate for ABA is $R(0, 0)$.  The different algorithms under comparison are assumed to deliver the same final output $\hat{p}$. 

Theorem \ref{thm3} presents an exact relation between the global rates for the same $r$ but different $f$; see Appendix D for its proof. 

\begin{theorem}
\label{thm3}
We have 
\begin{equation}
\label{rrf}
R(r, f) = (1+f_+) R(r, 0)-f_+.
\end{equation}
Consequently, $R(r, f)\leq R(r, \tilde{f})$ if $f_+\geq \tilde{f}_+$. 
\end{theorem} 

{\bf Remark.}  By writing $R(r, \tilde{f})$, we already assume that (\ref{tW}) is satisfied with $\tilde{f}$ in place of $f$.  See also Theorem \ref{thm4} below. 

For fixed $r$, Theorem \ref{thm3} simply recommends large values of $f_+$ for fast convergence.  In view of the constraint (\ref{W*}), this implies that, given $r$, $R(r, f)$ is minimized by 
\begin{equation}
\label{optf}
f_j=\frac{\min_i W^*_{ij} }{1-\sum_k \min_i W^*_{ik} },\quad 1\leq j\leq n, 
\end{equation}
where $W^*$ is defined in (\ref{rsat}).  This also leads to a nontrivial bound on $R(0, 0)$.
\begin{corollary}
\label{coro1}
$R(0, 0)\geq \sum_j \min_i W_{ij}.$
\end{corollary}
\begin{IEEEproof}
We have $R(r, 0)\geq f_+/(1+f_+)$ from (\ref{rrf}), since $R(r,f)\geq 0$.  The claim follows by choosing $r=0$ and $f$ as in (\ref{optf}). 
\end{IEEEproof}

Corollary \ref{coro1} formalizes the intuition that ABA is likely to be slow when there exists heavy overlap between rows of $W$.  The quantity $\sum_j \min_i W_{ij}$ is, in a sense, a conservative measure of this overlap. 

To compare the global rates for different values of $r$, it is convenient to write 
\begin{equation}
\label{g}
g=\frac{1+f_+}{1-r_+} rW+f.
\end{equation}
Then $f$ can be recovered from $g$ via
\begin{equation}
\label{fg}
f=g-(1+g_+) rW,\quad g_+=g 1_n.
\end{equation}
Let us define 
$$\tilde{R}(r, g)=R(r, f)$$
in view of this correspondence. 
\begin{corollary}
\label{trans}
For fixed $r,\ \tilde{R}(r, g)$ decreases in $g_+$. 
\end{corollary}
\begin{IEEEproof}
Noting 
\begin{equation}
\label{f+g+}
f_+=(1+g_+)(1-r_+)-1, 
\end{equation}
the claim follows from Theorem \ref{thm3}. 
\end{IEEEproof}

An advantage of using $g$ is that its optimal choice does not depend on $r$. 
\begin{proposition}
\label{prop5}
For fixed $r$ that satisfies (\ref{W}), $\tilde{R}(r, g)$ is minimized by 
\begin{equation}
\label{optg}
g_j=\frac{\min_i W_{ij}}{1-\sum_k \min_i W_{ik}},\quad 1\leq j\leq n.
\end{equation}
\end{proposition}
\begin{IEEEproof}
By direct calculation, (\ref{optg}) follows from (\ref{g}) and (\ref{optf}). 
\end{IEEEproof}

Theorem \ref{thm4} compares the global rates as a function of $r$ when $g$ is fixed.  The proof is presented in Appendix D.
\begin{theorem}
\label{thm4}
For fixed $g$, $\tilde{R}(r, g)$ decreases in $r/(1-r_+)$, i.e., 
$$\frac{r}{1-r_+}\geq \frac{\tilde{r}}{1-\tilde{r}_+}\quad \Longrightarrow \quad \tilde{R}(r, g)\leq \tilde{R}(\tilde{r}, g).$$ 
\end{theorem}

Theorem \ref{thm4} is relatively strong.  It implies Corollary \ref{weak}, as can be verified from Theorem \ref{thm3} and (\ref{g}).
\begin{corollary}
\label{weak}
For fixed $f,\ R(r, f)$ decreases in $r/(1-r_+)$.  Consequently $R(r, f)$ decreases in $r$.
\end{corollary}

Overall the function $\tilde{R}(r, g)$ decreases in both $r/(1-r_+)$ and $g$.  Since the original ABA corresponds to $(r, g)=(0, 0)$, Algorithm III is never worse than the original ABA in terms of the global rate. 

\begin{corollary}
\label{coro2}
We have 
$$R(r, f)\equiv \tilde{R}(r, g)\leq \tilde{R}(0, 0)\equiv R(0,0).$$
\end{corollary}

Theorem \ref{thm4} and Proposition \ref{prop5} lead to a general rule for choosing the ``squeezing parameters''. One should choose the largest allowable $g$ as specified by (\ref{optg}), and then choose a large $r/(1-r_+)$ subject to (\ref{W}).  For $m=2$ this resolves the optimal choice of $(r, g)$ completely.
\begin{corollary}
If $m=2$, then $\tilde{R}(r, g)$ is minimized when $g$ satisfies (\ref{optg}) and $r$ satisfies the equalities in (\ref{r1}) and (\ref{r2}). 
\end{corollary}

For general $m>2,$ finding the optimal $r$ appears nontrivial.  Fortunately, the optimal $r$ is not strictly necessary for achieving substantial improvements.  In Examples 1 and 2, Algorithm I, i.e., $r\equiv 0,$ is already considerably faster than ABA.  If the optimal $r$ is difficult to find, an option is to fix some $q\in \Omega$, and set $r=\delta q,\ \delta\geq 0$.  The constraint (\ref{W}) reduces to 
$$\delta\leq \min_{i, j} \frac{W_{ij}}{(qW)_j}.$$
Then we can set $\delta$ at this upper bound.  We leave the choice of $q$ as an open problem for further investigations. 

{\bf Remark.} Results in this section carry over to Algorithm II since Algorithm III is equivalent to Algorithm II with $\lambda$ given by (\ref{lamfr}).  By (\ref{lamfr}), for example, (\ref{f+g+}) simply says $1+g_+=\lambda$.  Hence Corollary \ref{trans} recommends setting $\lambda$ at its upper bound in (\ref{lam2}).  In view of (\ref{lamfr}), it is not surprising that in Theorem \ref{thm3} and Corollary \ref{trans}, the vectors $f$ and $g$ enter the picture only through $f_+$ and $g_+$. 

\section{Summary and discussion}
A simple ``squeezing'' strategy is studied for speeding up the Arimoto-Blahut algorithm for discrete memoryless channels.  This strategy  introduces auxiliary vectors $r$ and $f$ and reformulates the problem so as to reduce the overlap between rows of the channel matrix $W$.  A desirable feature of the resulting Algorithm II/III is that it improves ABA without sacrificing its simplicity or monotonic convergence properties. 

The effectiveness of Algorithm II/III is limited by the availability of large values of $r$ and $f$.  If the constraint (\ref{tW}) forces both $r$ and $f$ to be close to zero, then we can expect little improvement from Algorithm II/III.  Simply put, some channel matrices are not very ``squeezable.''  Nevertheless, modifications can conceivably be designed for such situations.  For example, suppose the input alphabet is ordered so that the overlap between conditional distributions $W_i$ is most severe between adjacent $i$'s.  Then a natural strategy is to apply Algorithm II to update the probabilities for one neighborhood of $i$'s at a time, holding the remaining components fixed.  Potential applications, e.g., to the discrete-time Poisson channel (\cite{S, LM}), will be reported in future works. 

An open problem is to determine the optimal squeezing parameters, i.e., the values of $r$ and $f$ that produce the fastest Algorithm III.  While the results in Section V paint a general picture, further theoretical studies may lead to extensions and refinements.  If the optimal choice is difficult to derive or to implement, empirical studies may suggest effective rules. 

\section*{Acknowledgments}
The author would like to thank Professors Donald Rubin, Xiao-Li Meng and David van Dyk for introducing him to statistical computing and related fields. 

\appendix
\section*{A: Waterfilling for (\ref{water})}
We need to determine $\delta\equiv \delta^{(t)}$ such that
$$\sum_i \max\{r_i,\, \delta x_i\}=1,$$
where $x_i=p_i^{(t)}\exp\left(\lambda z_i^{(t)}\right)$ as in (\ref{water}).  This is feasible with $\delta>0$ because $\sum_i r_i<1$. 

Step 1. Sort $r_i/x_i$, say 
$$\frac{r_1}{x_1}\leq \frac{r_2}{x_2}\leq\ldots\leq \frac{r_m}{x_m}.$$

Step 2. Calculate the cumulative sums $r^*_i=\sum_{j=i}^m r_j$ and $x_{i*}=\sum_{j=1}^i x_j,\ i=1,\ldots, m$.  By convention $r_{m+1}^*=x_{0*}=0$. 

Step 3. Locate the largest index $i\in\{1,\ldots, m\}$ such that 
$$\frac{r_i}{x_i} x_{i*} +r_{i+1}^*\leq 1.$$
Set $\delta =(1-r^*_{i+1})/x_{i*}.$

The overall time cost is $O(m\log m)$ due to Step 1. 

\section*{B: Proof of Theorem \ref{thm1}: monotonic convergence}
Algorithm III is seen as an alternating divergence minimization procedure between convex sets of measures (\cite{CT, CS}).  Let $X=\{0, 1, \ldots, m\}$ and $Y=\{1,\ldots, n\}$.  Let 
$\mathcal{P}$ be the set of measures on $X\times Y$ of the form $P=(P_{ij}),$ 
$$P_{ij}=\left\{\begin{array}{cc} p_i\tilde{W}_{ij}, & 1\leq i\leq m\\ f_j, & i=0\end{array}\right. $$
where $p=(p_1,\ldots, p_m)\in \Omega(r)$.  Let $\mathcal{Q}$ be the set of measures on $X\times Y$ of the form $Q=(Q_{ij}),$
$$Q_{ij}=\left\{\begin{array}{cc} \Phi_{ji}\tilde{W}_{ij}e^{c_i}, & 1\leq i\leq m\\ f_j \Phi_{j0}, & i=0\end{array}\right.$$
where $\Phi_{ji}\geq 0$ and $\sum_{i=0}^m \Phi_{ji}=1$.  Observe that (i) both $\mathcal{P}$ and $\mathcal{Q}$ are convex; (ii) $I(p, \Phi)=-D(P||Q);$ and (iii) (\ref{maxphi}) and (\ref{maxp1}) correspond to minimizing $D(P||Q)$ over $Q$ for fixed $P$, and over $P$ for fixed $Q$, respectively.  The claim then follows from Theorem 3 of Csisz\'{a}r and Tusnady \cite{CT}. 

\section*{C: Proof of Theorem \ref{thm2}: convergence rate}
With a slight abuse of notation let $\Phi_{ji}(p)$ and $p_i(\Phi)$ be functions given by (\ref{maxphi}) and (\ref{maxp}) respectively.  Then ($1\leq i, k\leq m,\ 1\leq j\leq n$)
\begin{equation}
\label{deri_phi}
\frac{\partial \Phi_{ji}(p)}{\partial p_k}=\left\{
\begin{array}{ll} \Phi_{ji}(p) (1-\Phi_{ji}(p)) p_i^{-1}, & k=i;\\
-\Phi_{ji}(p) \Phi_{jk}(p) p_k^{-1}, & k\neq i.\end{array}\right.
\end{equation}
\begin{equation}
\label{deri_p}
\frac{\partial p_{i}(\Phi)}{\partial \Phi_{jk}}=\left\{
\begin{array}{ll} p_i(\Phi)(1-p_i(\Phi)) \tilde{W}_{ij} \Phi_{ji}^{-1}, & k=i;\\
-p_{i}(\Phi) p_k(\Phi) \tilde{W}_{kj} \Phi_{jk}^{-1}, & k\neq i.\end{array}\right.
\end{equation}
We calculate $R(p^*)$ as 
$$R(p^*)=\left. \frac{\partial p(\Phi(p))}{\partial p}\right|_{p=p^*}.$$
Write $\Phi^*=\Phi(p^*)$.  Then $p(\Phi^*)=p^*$.  These relations and (\ref{deri_phi}) and (\ref{deri_p}) are used repeatedly. 

For $i\neq k,\ 1\leq i,k\leq m$, we have 
\begin{align}
\nonumber
\frac{\partial p_i(\Phi(p^*))}{\partial p_k} = &\sum_j \left[\frac{\partial p_i(\Phi^*)}{\partial \Phi_{ji}} \frac{\partial \Phi_{ji}(p^*)}{\partial p_k} + \frac{\partial p_i(\Phi^*)}{\partial \Phi_{jk}} \frac{\partial \Phi_{jk}(p^*)}{\partial p_k}\right]\\
\nonumber
&+\sum_j \sum_{l\geq 1,\, l\neq i,\, l\neq k} \frac{\partial p_i(\Phi^*)}{\partial \Phi_{jl}} \frac{\partial \Phi_{jl}(p^*)}{\partial p_k}\\
\nonumber
=& -\sum_j p_i^*\left[\frac{1-p_i^*}{p_k^*} \tilde{W}_{ij} \Phi_{jk}^* + \tilde{W}_{kj} (1-\Phi_{jk}^*)\right]\\ 
\nonumber
&+ \sum_j \sum_{l\geq 1,\, l\neq i,\, l\neq k} \frac{p_i^*p_l^*}{p_k^*} \tilde{W}_{lj} \Phi_{jk}^*\\
\label{s1}
=& -\sum_j \left[(1-p_i^*) \tilde{W}_{kj} \Phi_{ji}^* + p_i^* \tilde{W}_{kj} (1-\Phi_{jk}^*)\right]\\ 
\nonumber
&+ \sum_j p_i^*(1-\Phi_{j0}^*-\Phi_{ji}^*-\Phi_{jk}^*) \tilde{W}_{kj}\\
\label{ik}
=&-\sum_j \tilde{W}_{kj}(\Phi^*_{ji}+p_i^*\Phi_{j0}^*),
\end{align}
where (\ref{s1}) uses (\ref{maxphi}). 

For $1\leq k\leq m$, a similar calculation yields 
\begin{equation}
\label{kk}
\frac{\partial p_k(\Phi(p^*))}{\partial p_k} = 1-\sum_j \tilde{W}_{kj}(\Phi^*_{jk}+p_k^*\Phi_{j0}^*).
\end{equation}
Alternatively, (\ref{kk}) can be derived from (\ref{ik}) and 
\begin{equation}
\label{sum0}
\sum_i \frac{\partial p_i(\Phi(p^*))}{\partial p_k}=\frac{\partial \sum_i p_i(\Phi(p^*))}{\partial p_k}=0.
\end{equation}
The identity (\ref{key}) is just (\ref{ik}) and (\ref{kk}) in matrix format. 

\section*{D: Convergence rates: properties and comparisons}
This section proves Propositions \ref{prop2} and \ref{prop3}, and Theorems \ref{thm3} and \ref{thm4}.  The notation is the same as in Section V. 

Part 1 of Proposition \ref{prop2} follows from (\ref{sum0}).  For further analysis, define 
$$W^*=\frac{I_m- 1_m r}{1-r_+}W,\quad s=p^*W^*,\quad D_s={\rm Diag(s)}.$$ 
That is, $D_s$ is the diagonal matrix with $s$ as the diagonal entries.  Also let $D_{p^*}={\rm Diag}(p^*)$.  From (\ref{maxphi}) and (\ref{Psi}), we obtain 
$$\Psi=D_s^{-1} W^{*\top} D_{p^*}.$$
Thus (\ref{key}) can be written as 
\begin{equation}
\label{rp*}
R(p^*) =I_m-(1+f_+)K+L
\end{equation}
where 
\begin{equation*}
K =W^* D_s^{-1}W^{*\top} D_{p^*};\quad L =1_m f D_s^{-1}W^{*\top} D_{p^*}.
\end{equation*}
Observe that $D_{p^*}^{1/2} K D_{p^*}^{-1/2}$ is symmetric and nonnegative definite.  Thus $K$ is diagonalizable and has only nonnegative eigenvalues.  When $f\equiv 0$, we have $R(p^*)= I_m - K$.  Thus $R(p^*)$ is diagonalizable in this case.  This proves Proposition \ref{prop2}. 

Define a space of row vectors $\Gamma=\{\gamma\in \mathbf{R}^m:\ \gamma 1_m=0\}$.  For an $m\times m$ matrix $A$ such that $\gamma A\in \Gamma$ whenever $\gamma\in \Gamma$, we write $S_0(A)$ as the spectral radius of $A$ when restricted as a linear transformation on $\Gamma$.  Suppose $A$ satisfies $A 1_m=0$, and suppose $d$ is a nonzero eigenvalue of $A$, with a corresponding left eigenvector $\gamma$.  Then 
$$0=\gamma A 1_m=d\gamma 1_m\ \ \Longrightarrow\ \ \gamma \in \Gamma.$$
Hence the set of nonzero eigenvalues is unchanged when $A$ is restricted to $\Gamma$.  In particular, 
\begin{equation}
\label{eqS}
S(A)=S_0(A).
\end{equation}

We have $\gamma L=0$ for any $\gamma\in \Gamma$.  Thus $R(p^*)$ and $I_m-(1+f_+)K$ represent the same linear transformation when restricted to $\Gamma$.  Also, $R(p^*) 1_m=0$ by Proposition \ref{prop1}.  By the preceding discussion, if $d$ is a nonzero eigenvalue of $R(p^*)$, then $d$ is an eigenvalue of $I_m-(1+f_+)K$.  Equivalently, $(1-d)/(1+f_+)$ is an eigenvalue of $K$.  We know $d\leq 1$ because $K$ only has nonnegative eigenvalues.  On the other hand, because $1-d$ is an eigenvalue of the stochastic matrix $\tilde{W}\Psi$, the Frobenius-Perron theorem implies that $|1-d|\leq 1$, i.e., $d\geq 0$.  This proves Proposition \ref{prop3}. 

We also have 
\begin{align}
\label{RS}
R(r, f) &=S_0(R(p^*))\\
\nonumber
        &=S_0(I_m-(1+f_+)K)\\
\label{S0}
        &=(1+f_+)S_0(I_m-K)-f_+\\
\label{SK}
        &=(1+f_+)S(I_m-K)-f_+\\
\label{Rr0}
        &=(1+f_+)R(r, 0)-f_+. 
\end{align}
Identity (\ref{RS}) follows from (\ref{eqS}).  Identity (\ref{S0}) holds because, by Proposition \ref{prop3}, the spectral radii involved refer to the largest eigenvalues.  Because $(I_m-K) 1_m=0$, we have 
(\ref{SK}).  Identity (\ref{Rr0}) holds because $I_m-K$ is precisely the matrix rate of Algorithm III that uses $(r, 0)$ in place of $(r, f)$.  Thus we have proved Theorem \ref{thm3}. 

\begin{IEEEproof}[Proof of Theorem \ref{thm4}]
By (\ref{fg}) and Theorem \ref{thm3}, we have
\begin{align*}
1-\tilde{R}(r, g) &=1-R(r,\, g-(1+g_+)rW)\\
&=(1+g_+)(1-r_+) (1-R(r, 0)).
\end{align*}
Thus, to prove $\tilde{R}(r, g)\leq \tilde{R}(\tilde{r}, g),$ we only need 
$$(1-r_+)(1-R(r, 0))\geq (1-\tilde{r}_+)(1-R(\tilde{r},0)).$$
Let us only consider $\tilde{r}\equiv 0$, i.e., 
\begin{equation}
\label{onlyneed}
(1-r_+)(1-R(r, 0))\geq 1-R(0,0).
\end{equation}
The general case reduces to this special one (details omitted) if we replace $W$ by 
$$\frac{I_m-1_m \tilde{r}}{1-\tilde{r}_+} W,$$
and $r$ by $r-(1-r_+) \tilde{r}/(1-\tilde{r}_+)$. 

By (\ref{rp*}), we have 
$$R(r,0) =S(I_m-U F U^\top D_{p^*})$$
where 
$$U=\frac{I_m- 1_m r}{1-r_+},\quad F =WD_s^{-1}W^\top,$$
$$s=p^*UW=\hat{p} W,\quad p^*=(1-r_+)\hat{p}+r,$$ 
and $\hat{p}$ denotes the (same) final output of Algorithm III using $(r, 0)$ or $(0, 0)$ for $(r, f)$.  Define 
\begin{equation}
\label{adef}
A =F U^\top D_{p^*}.
\end{equation} 
The same argument leading to Proposition \ref{prop3} and Theorem \ref{thm3} shows that all eigenvalues of $A$ are in the interval $[0,1]$, and 
\begin{equation}
\label{speca}
S(I_m-A) =(1-r_+)R(r, 0)+r_+.
\end{equation}
Define 
\begin{align}
\nonumber
C &\equiv F^{1/2} r^\top \hat{p} F^{1/2};\\
\label{atilde}
\tilde{A} &\equiv F^{1/2} U^\top D_{p^*} F^{1/2}\\
\nonumber
&=F^{1/2} \left(D_{\hat{p}}+\frac{D_r-r^\top r}{1-r_+}\right)F^{1/2}-C. 
\end{align}
Comparing (\ref{atilde}) with (\ref{adef}) shows that $\tilde{A}$ and $A$ have the same set of eigenvalues.  Let $a$ be the smallest eigenvalue of $\tilde{A}$, and let $\beta$ be a corresponding right eigenvector.  Then $a=1-S(I_m-\tilde{A})$, and by (\ref{speca}), 
\begin{equation}
\label{aless}
a =1-S(I_m-A)=(1-r_+)(1-R(r, 0)).
\end{equation}
By direct calculation, we have 
\begin{equation}
\label{betaC}
aC\beta=C\tilde{A}\beta=[(1-r_+) C +F^{1/2} r^\top r F^{1/2}]\beta.
\end{equation}
If $a=1-r_+$, then (\ref{betaC}) gives $F^{1/2} r^\top r F^{1/2}\beta=0,$ which implies 
$$\beta^\top F^{1/2} r^\top r F^{1/2}\beta=0;\quad r F^{1/2}\beta=0;\quad \beta^\top C=0.$$
Thus, 
\begin{align}
\label{a_s1}
a\beta^\top\beta &=\beta^\top \tilde{A}\beta \\
\nonumber
&=\beta^\top F^{1/2} \left(D_{\hat{p}}+\frac{D_r}{1-r_+}\right) F^{1/2}\beta\\
\nonumber
&\geq \beta^\top F^{1/2} D_{\hat{p}} F^{1/2}\beta\\
\label{a_s4}
&\geq (1-R(0, 0))\beta^\top \beta,
\end{align}
where (\ref{a_s4}) follows from 
$$R(0,0)=S(I_m-FD_{\hat{p}})=S(I_m-F^{1/2} D_{\hat{p}} F^{1/2}).$$
We deduce $a\geq 1-R(0, 0)$ and conclude the proof of (\ref{onlyneed}).  If $a\neq 1-r_+$, then (\ref{aless}) implies $a+r_+-1<0$, and 
(\ref{betaC}) leads to 
$$\beta^\top C\beta =\frac{\beta^\top F^{1/2} r^\top r F^{1/2}\beta}{a+r_+-1}.$$
Calculations similar to (\ref{a_s1})--(\ref{a_s4}) yield the same conclusion, i.e., 
$a\geq 1-R(0, 0)$. 
\end{IEEEproof}



\begin{thebibliography}{10}
\bibitem{A}
S. Arimoto, ``An algorithm for computing the capacity of arbitrary
discrete memoryless channels,'' {\it IEEE Trans. Inform. Theory}, vol. 18, pp. 14--20, 1972. 
\bibitem{B}
R. E. Blahut, ``Computation of channel capacity and rate-distortion
functions,'' {\it IEEE Trans. Inform. Theory}, vol. 18, pp. 460--473, 1972.
\bibitem{Cover}
T. Cover and J. Thomas, {\it Elements of Information Theory}, 2nd ed., New York: Wiley, 2006.
\bibitem{CS}
I. Csisz\'{a}r and P. Shields, ``Information theory and statistics: a tutorial,'' {\it Foundations and Trends in 
Communications and Information Theory,} vol. 1, pp. 417--528, 2004. 
\bibitem{CT}
I. Csisz\'{a}r and G. Tusnady, ``Information geometry and alternating minimization procedures,''
{\it Statistics \& Decisions} Supplement Issue 1, pp. 205--237, 1984.
\bibitem{DLR}
A. P. Dempster, N. M. Laird and D. B. Rubin, ``Maximum likelihood estimation from incomplete
data via the EM algorithm'' (with discussion), {\it J. Roy. Statist. Soc. B}, vol. 39, pp. 1--38, 1977.
\bibitem{DYW}
F. Dupuis, W. Yu and F.M.J. Willems, ``Blahut-Arimoto algorithms for computing channel capacity and rate-distortion with side information,'' {\it Proc. 2004 International Symposium on Information Theory}, Chicago, IL, June/July 2004. 

\bibitem{G}
R. G. Gallager, {\it Information Theory and Reliable Communication,} Wiley: New York, 1968. 

\bibitem{LM}
A. Lapidoth and S. M. Moser, ``On the capacity of the discrete-time Poisson channel,'' {\it IEEE Trans. Inf. Theory},
vol. 55, no. 1, pp. 303--322, 2009.

\bibitem{MD}
G. Matz and P. Duhamel, ``Information geometric formulation and interpretation of accelerated Blahut-Arimoto-type algorithms,'' In {\it Proc. 2004 Information Theory Workshop}, Oct. 2004. 

\bibitem{XLM}
X. L. Meng, ``On the rate of convergence of the ECM algorithm,'' {\it Ann. Statist.}, vol. 22, no. 1, pp. 326--339, 1994. 

\bibitem{N}
H. Nagaoka, ``Algorithms of Arimoto-Blahut type for computing
quantum channel capacity,'' In {\it Proc. 1998 International Symposium on Information Theory}, Cambridge, MA, Aug. 1998.
\bibitem{RG}
M. Rezaeian and A. Grant, ``A generalization of the Arimoto-Blahut algorithm,'' {\it Proc. 2004 International Symposium on Information Theory}, Chicago, IL, June/July 2004.

\bibitem{S}
S. Shamai (Shitz), ``Capacity of a pulse amplitude modulated direct
detection photon channel,'' in {\it Proc. Inst. Elec. Eng.}, vol. 137, no. 6, pp.
424–-430, Dec. 1990, part I (Communications, Speech and Vision).

\bibitem{V}
P. 0. Vontobel, ``A generalized Blahut-Arimoto algorithm,'' In {\it Proc. 2003 International Symposium on Information Theory}, Yokohama, Japan, June/July 2003. 
\end{thebibliography}
\end{document}